\documentclass{article}
\usepackage{spconf,amsmath,amssymb,bm,graphicx}
\usepackage[hidelinks]{hyperref}
\usepackage[capitalize,nameinlink]{cleveref}
\usepackage{subcaption}
\usepackage{booktabs}
\usepackage[export]{adjustbox}
\usepackage{microtype}
\usepackage{flushend}
\usepackage[backend=biber, bibencoding=utf8,
            style=ieee,
            maxbibnames=6,
            doi=false,
            url=false,
            isbn=false]{biblatex}
\AtEveryBibitem{%
  \ifentrytype{inproceedings}{%
    \clearfield{pages}%
    \clearlist{publisher}%
    \clearlist{organization}%
    \clearlist{location}}{}%
  \ifentrytype{misc}{%
    \clearfield{eprintclass}{} }}

\title{Are These Even Words? Quantifying the Gibberishness\\of Generative Speech Models}

\name{Danilo de Oliveira, Tal Peer, Jonas Rochdi, Timo Gerkmann}
\address{Signal Processing, University of Hamburg, Germany}

\usepackage[nolist]{acronym}

\begin{acronym}
\acro{ASR}{automatic speech recognition}
\acro{DNN}{deep neural network}
\acro{PER}{phoneme error rate}
\acro{WER}{word error rate}
\acro{SNR}{signal-to-noise ratio}
\acro{STFT}{short-time Fourier transform}
\acro{PESQ}{perceptual evaluation of speech quality}
\acro{ESTOI}{extended short-time objective intelligibility}
\acro{DNSMOS}{deep noise suppression mean opinion score}
\acro{MOS}{mean opinion score}
\acro{LPS}{Levenshtein phoneme similarity}
\acro{CTC}{connectionist temporal classification}
\acro{SSL}{self-supervised learning}
\acro{EEG}{electroencephalography}
\acro{FFT}{fast Fourier transform}
\acro{GPU}{graphics processing unit}
\acro{ROI}{region-of-interest}
\acro{EMA}{exponential moving average}
\acro{DAC}{descript audio codec}
\acro{VSR}{visual speech recognition}
\acro{EMA}{exponential moving average}
\acro{LM}{language model}
\acro{LLM}{large language model}
\acro{uLM}{unit language model}
\acro{SpeechLM}{speech language model}
\acro{VQ-VAE}{vector-quantized variational autoencoder}
\acro{SE}{speech enhancement}
\acro{PCC}{Pearson's correlation coefficient}
\acro{SRCC}{Spearman's rank correlation coefficient}
\acro{L2S}{lip-to-speech}
\end{acronym}

\begin{document}
\maketitle
\begin{abstract}
    Significant research efforts are currently being dedicated to non-intrusive quality and intelligibility assessment, especially given how it enables curation of large scale datasets of in-the-wild speech data. However, with the increasing capabilities of generative models to synthesize high quality speech, new types of artifacts become relevant, such as generative hallucinations. While intrusive metrics are able to spot such sort of discrepancies from a reference signal, it is not clear how current non-intrusive methods react to high-quality phoneme confusions or, more extremely, gibberish speech. In this paper we explore how to factor in this aspect under a fully unsupervised setting by leveraging language models. Additionally, we publish a dataset of high-quality synthesized gibberish speech for further development of measures to assess implausible sentences in spoken language, alongside code for calculating scores from a variety of speech language models.\footnote{\url{https://uhh.de/inf-sp-gibberish}}
\end{abstract}
\section{Introduction}
\vspace{-8pt}
\label{sec:intro}
Deep generative models have been recently introduced into many fields, complementing and sometimes replacing existing predictive approaches. In particular, the field of speech signal processing has been experiencing a drastic increase in the number of generative models being introduced into the research landscape and also being adopted in various applications. Generative speech models are able to produce natural sounding utterances that belong to a modeled speech distribution~\cite{richter2023speech}. This greatly benefits tasks involving speech generation, such as text-to-speech, \ac{SE}, lip-to-speech and speech language modeling. 

While the advance in generative modeling allows for groundbreaking improvement in performance, it also introduces new challenges, including the question of proper, fair and insightful evaluation. Generative models are able to produce speech signals of very high quality, but are susceptible to a class of distortions which is virtually non-existent for predictive models: \emph{hallucinations}. Hallucinations in generative speech models can appear in several different forms, including phonetic confusions or filling-in of silent parts with speech-like or paralinguistic sounds (e.g. gasps or sighs)~\cite{lemercier2023analysing}. The most extreme manifestation of such hallucinations is the generation of completely incomprehensible words or passages which are composed of speech-like sounds but lack any semantic meaning; this kind of generated speech will be referred to in this paper as ``gibberish''. This failure mode of generative speech models tends to emerge especially when the  input to the model is severely degraded or ambiguous. Examples include \ac{SE} on low \ac{SNR} input~\cite{lemercier2023storm}, or the lip-to-speech task, where speech is synthesized only from lip movement without an audio signal~\cite{deoliveira2025lipdiffuser}.

\begin{figure}[t]
    \centering
    \begin{subfigure}{.48\columnwidth}
            \centering
            \includegraphics[width=\linewidth]{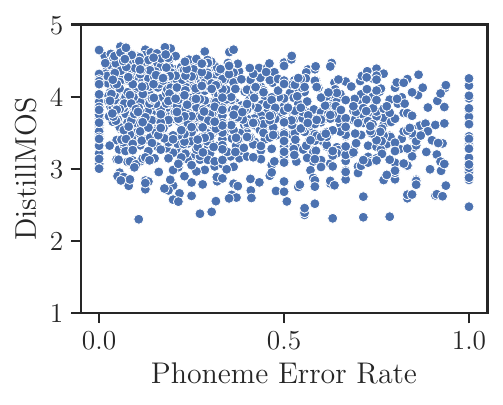}
            \caption{}
            \label{fig:lipdiffuser_per}
    \end{subfigure}
    \hfill
    \begin{subfigure}{.48\columnwidth}
            \centering
            \includegraphics[width=\linewidth]{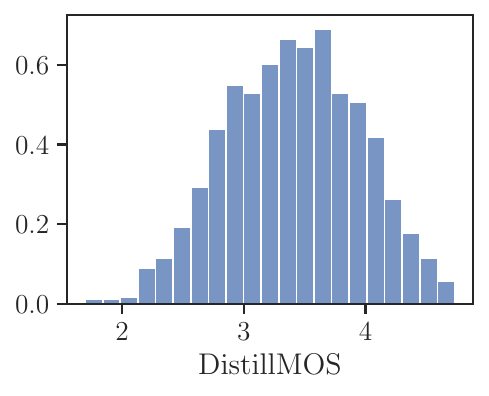}
            \caption{}
            \label{fig:gibberish_utmosv2}
    \end{subfigure}
    \caption{\subref{fig:lipdiffuser_per} DistillMOS vs. PER evaluated on LipDiffuser outputs. \subref{fig:gibberish_utmosv2} Normalized histogram of DistillMOS scores on gibberish speech generated by an unconditional diffusion model. Details on the data for both plots are given in \cref{sec:data}.}
    \label{fig:intro_figs}
\end{figure}
The evaluation of speech signals and systems is an actively researched topic. Many evaluation methods exist, aiming to quantify different aspects such as quality, intelligibility, noisiness, etc. Since speech is such an integral part of the human experience, perceptual (``subjective'') studies involving human subjects remain the gold standard for evaluation. However, these studies are costly and time-intensive, necessitating the development of instrumental metrics that, ideally, correlate well to the human-based results. Instrumental metrics can be categorized into intrusive metrics, which compare a test signal with a reference signal, and non-intrusive (blind) metrics, which do not require a reference signal. 

Non-intrusive metrics allow evaluation on large unlabeled datasets and are becoming more popular recently, especially such metrics based on \acp{DNN}. In the context of generative models, it has been shown that both intrusive and non-intrusive metrics have their own advantages and drawbacks~\cite{lemercier2023analysing,deoliveira2023behavior,pirklbauer2023evaluation}. In particular, intrusive metrics heavily penalize generative models due to hallucinations, despite very high perceived quality. At the same time, non-intrusive metrics are able to reflect this high perceived quality, but are generally unable to capture the loss of linguistic content caused by hallucinations.

In order to further analyze this behavior, we first make a distinction between two types of intrusive metrics. \emph{Signal-intrusive} metrics measure the acoustic similarity between two signals, directly based on the signals' energy (e.g. SI-SDR) or by employing an auditory model (e.g. PESQ, POLQA). On the other hand, \emph{content-intrusive} metrics such as \ac{WER} or \ac{PER} do not measure whether two signals sound the same. Instead, they reveal whether the two signals convey the same message, and to what extent. This is done by transcribing the audio signals into either words or phonemes using an \ac{ASR} system. Note that while some intrusive metrics such as STOI or HASPI specifically aim to predict intelligibility, these metrics are still computed at the acoustic signal level~\cite{vankuykEvaluationIntrusiveInstrumental2018} and we thus treat them as signal-intrusive.

Although not explicitly signal-oriented, existing non-intrusive speech metrics usually correlate well to signal-intrusive metrics. The same holds for content-intrusive metrics considering predictive models~\cite{deoliveira2023behavior}. However, in the case of generative models we observe a discrepancy w.r.t. content-intrusive metrics, due to hallucinations. This effect is demonstrated in \cref{fig:lipdiffuser_per}, which shows an evaluation of a recent generative lip-to-speech system~\cite{deoliveira2025lipdiffuser}. The content-intrusive \ac{PER} metric can measure the loss of information caused by hallucinations, but this loss is not captured by the non-intrusive DistillMOS metric~\cite{stahl2025distillation}, which is ``fooled'' by the consistently high acoustic quality achieved by the lip-to-speech model. An even more extreme example is shown in \cref{fig:gibberish_utmosv2} where complete gibberish speech generated by an unconditional diffusion model consistently scores well on DistillMOS.

In search of a non-intrusive metric which can capture the loss or preservation of linguistic content, we hypothesize that the inclusion of a \ac{LM} can assist in this task by providing linguistic cues that can help discern linguistically plausible signals from gibberish. A few contributions considering the use of \acp{LM} for non-intrusive speech signal evaluation have already been proposed. In~\cite{zezario2025study} a simple approach is introduced, where the \ac{LLM} GPT-4o is presented with the audio input and prompted to rate its quality and intelligibility. A more nuanced approach using a \ac{SpeechLM}, explicitly trained on tokenized audio input, is described in~\cite{maiti2023speechlmscore}. In this case the metric (SpeechLMScore) is computed from the log-likelihood of the predicted token sequence (perplexity, see~\cref{sec:perplexity}). While the authors of~\cite{maiti2023speechlmscore} report high correlation with human perceptual scores, they do not analyze whether the \ac{SpeechLM}-based score reacts differently to acoustic or linguistic degradations. 

In this paper, we expand upon the SpeechLMScore approach, and demonstrate how it can be modified to better capture linguistic aspects. Furthermore, we extend this approach to other \acp{SpeechLM} and perform an extensive comparison of their ability to assess phonetic confusions and gibberishness in audio. In order to promote further research in this field, we also publish a dataset of high-quality gibberish speech, enabling evaluation of future non-intrusive metrics' sensitiveness to gibberish.

\section{Method}
\vspace{-5pt}
\label{sec:method}
\subsection{Autoregressive Speech Language Models}
\vspace{-5pt}
Autoregressive \acp{SpeechLM} represent audio as sequences of tokens, enabling the direct application of language modeling techniques. Usually, \ac{SSL} representations, such as HuBERT~\cite{hsu2021hubert} features, are quantized into a finite vocabulary using k-means clustering, yielding \textit{semantic tokens}. This approach, pioneered by GSLM~\cite{lakhotia2021generative}, allows autoregressive models to capture linguistic structure directly from speech. Alternative tokenization methods include residual vector quantization in \acp{VQ-VAE} and use of acoustic features to improve naturalness~\cite{defossez2024moshi, borsos2023audiolm}. 

In all cases, the model predicts each token given its history. Given a token sequence $\bm{x} = (x_1, \ldots, x_T)$, it estimates conditional probabilities $p_\theta(x_t \mid x_{<t})$ of each token $x_t$, where $\theta$ represents the model parameters. The joint distribution $p_\theta(\bm{x}) = \prod_t p_\theta(x_t \mid x_{<t})$ reflects how plausible a sequence is given the training data. 
The model is trained to minimize the cross-entropy loss

\begin{equation}
\mathcal{L}_{\text{CE}}(\bm{x}) = - \tfrac{1}{T} \sum_{t=1}^{T} \log p_\theta(x_t \mid x_{<t}).
\end{equation}

\subsection{Perplexity}\label{sec:perplexity}
\vspace{-5pt}
Perplexity is a standard evaluation metric in language modeling that quantifies how well a model predicts a sequence. It is directly related to the cross-entropy loss: for a sequence $\bm{x}$, the average cross-entropy per token is $\mathcal{L}_{\text{CE}}(\bm{x})$, and perplexity is defined as its exponential.

Inspired by~\cite{maiti2023speechlmscore}, we use cross-entropy (i.e.: log-perplexity) as a non-intrusive metric for assessing generated speech. The idea is that a well-trained \ac{SpeechLM} assigns higher probabilities to token sequences that resemble natural speech, and lower probabilities to sequences that contain unlikely or unnatural patterns. Perplexity directly reflects this likelihood: lower values indicate that the generated sequence is statistically closer to real speech in the training distribution, while higher values suggest that the sequence deviates from expected linguistic or acoustic structures. In this sense, perplexity can be interpreted as a way to quantify how natural a sequence of tokens appears under the model. However, in contrast to~\cite{maiti2023speechlmscore}, we separate the aspects of acoustic naturalness and linguistic plausibility, and specifically assess how a \ac{SpeechLM}'s log-perplexity reflects the latter.  %

\subsection{Metrics}\label{sec:implementation}
\vspace{-5pt}
We adapt several \acp{LM} to act as non-intrusive metrics:

\noindent\textbf{SpeechLMScore}~\cite{maiti2023speechlmscore} follows the GSLM framework, but replaces the large transformer with a lightweight LSTM-based \ac{uLM}. It deduplicates semantic tokens before modeling. The authors show good correlation with speech quality and naturalness scores and indicate a preference for using discrete units from layer 3 of HuBERT. However, given the many works showing a higher correlation of deeper layers with linguistic/semantic content~\cite{pasad2021layerwise, deoliveira2023leveraging}, we also include the results for layer 6. We denote these two as SpeechLMScore (3) and (6), respectively.

\noindent\textbf{VAE-GSLM}~\cite{chen2025variational} extends GSLM with a variational autoencoder branch that models continuous acoustic representations alongside semantic tokens. The \ac{LM} is trained jointly on both branches to capture complementary information. As in the original paper, we use the logits from the discrete branch for the computation of perplexity.

\noindent\textbf{TWIST}~\cite{hassid2023textually} builds upon the GSLM pipeline but initializes the \ac{uLM} from a pretrained \ac{LLM}. This warm-start allows TWIST to achieve better performance than a \ac{SpeechLM} trained from scratch. We use the version with 350M parameters.

\noindent\textbf{SpeechGPT}~\cite{zhang2023speechgpt} is an \ac{LLM} designed to handle both speech and text in a unified way. It is trained with discrete speech representations and a specialized instruction dataset, enabling it to follow cross-modal instructions and generate multi-modal content. We use the model from the first training stage, trained on speech continuation.

\noindent\textbf{\ac{ASR} + \ac{LLM}} is inspired by the semantic evaluation in Spectron~\cite{nachmani2024spoken}. We compare the use of Parakeet~\cite{rekesh2023fast} and QuartzNet~\cite{kriman2020quartznet} to transcribe audio, and a pretrained GPT-2~\cite{radford2019language} to compute the perplexity of the transcribed text.

\subsection{Analysis}\vspace{-5pt}
We generally compare the aforementioned \ac{LM}–based metrics with the \ac{PER} from a \ac{SSL}-based phoneme classifier~\cite{xu2022simple}, computed against the predictions for the original test set as a reference. Additionally, we employ supervised \ac{MOS} and intelligibility prediction metrics:

\noindent\textbf{UTMOSv2}~\cite{baba2024utmosv2} takes raw audio as input and directly predicts a \ac{MOS} that correlates with human judgments of naturalness. The model builds on self-supervised speech representations and is trained on large annotated corpora.

\noindent\textbf{DistillMOS}~\cite{stahl2025distillation} is a \ac{MOS} prediction approach that uses knowledge distillation. A teacher model trained on large-scale subjective ratings guides a lighter student model to predict quality scores efficiently.

\noindent\textbf{TorchAudio-Squim}~\cite{kumar2023torchaudiosquim} is a suite of models trained in a supervised manner to predict intrusive objective metrics in a non-intrusive fashion. We make use of the model which predicts STOI for intelligibility.

\section{Data}\label{sec:data}
\vspace{-8pt}
We test these models on multiple variants of the test set of LRS3~\cite{afouras2018lrs3}, covering clean, noisy, and generated speech conditions. LRS3 is a multi-modal dataset for audio-visual speech recognition, containing over 400 hours of video with matching audio of TED and TEDx talks, and its test set contains 1321 files, totaling one hour. 
For noisy conditions, we use LRS3-CHiME3~\cite{deoliveira2025lipdiffuser}, created by adding CHiME3 noise at \(-10\), \(-5\), \(0\), and \(5\) dB \ac{SNR} to LRS3. 

For the creation of generated speech, we use LipDiffuser~\cite{deoliveira2025lipdiffuser}, a conditional diffusion model that generates high-quality speech from silent video input. Its outputs are of high-quality sound, but can contain phonetic confusions or hallucinations in cases of low video quality, lip occlusion or lack of articulation. To take this to the extreme, we also extract the unconditioned outputs from the checkpoint of LipDiffuser's audio-pretraining stage. This model is conditioned only on speaker embeddings without any phonetic guidance, yielding high-quality gibberish speech devoid of semantic meaning. We also publish a dataset designed for testing non-intrusive metrics for sensitivity to gibberish speech, based on speaker characteristics learned from publicly available speech data. It can be seen a more extreme complement of the sWUGGY and sBLIMP benchmarks \cite{dunbar2021zeroresource} for evaluation of \acp{SpeechLM}.

\begin{table}
    \centering
    \caption{Correlations with \ac{PER} on noisy LRS3 + CHiME data at varying \acp{SNR}, and on lip-to-speech data, generated by LipDiffuser from silent videos.}
    \adjustbox{max width=\linewidth}{
    \begin{tabular}{lcccc}
        \toprule
         & \multicolumn{2}{c}{Noisy } & \multicolumn{2}{c}{Lip-to-speech}\\
         \cmidrule(lr){2-3}\cmidrule(lr){4-5}
         & $|\mathrm{PCC}|$ & $|\mathrm{SRCC}|$ & $|\mathrm{PCC}|$ & $|\mathrm{SRCC}|$ \\
        \midrule
        SpeechLMScore (3) & 0.532          & 0.505          & 0.402          & 0.414\\
        SpeechLMScore (6) & 0.691          & 0.694          & 0.545          & 0.558\\
        VAE-GSLM          & 0.401          & 0.345          & 0.624          & 0.676\\
        TWIST             & 0.625          & 0.586          & \textbf{0.705} & \textbf{0.707}\\
        SpeechGPT         & 0.767          & 0.736          & 0.592          & 0.594\\
        QuartzNet + GPT-2  & 0.452          & 0.452          & 0.477          & 0.500\\
        Parakeet + GPT-2   & 0.136          & 0.128          & 0.378          & 0.396\\
        \midrule
        UTMOSv2           & 0.189          & 0.202          & 0.293          & 0.261\\
        DistillMOS        & 0.772          & 0.753          & 0.306          & 0.301\\
        Squim STOI        & \textbf{0.806} & \textbf{0.828} & 0.101          & 0.111\\
        \bottomrule
    \end{tabular}}
    \label{tab:correlation}
\end{table}

\begin{figure*}
    \centering
    \includegraphics[width=\linewidth]{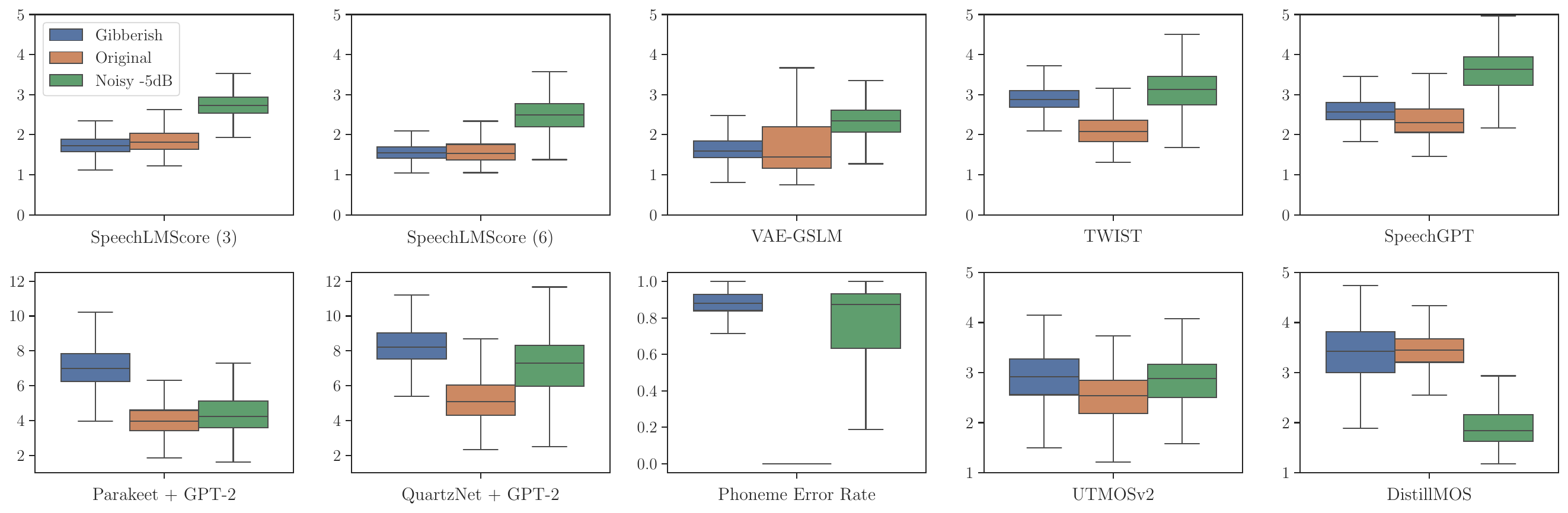}
    \caption{Estimated distributions of \ac{SpeechLM}-based metrics on gibberish, clean and noisy (-5dB \ac{SNR}) data. For UTMOSv2 and DistillMOS, higher values are better. For all others, lower is better.}
    \label{fig:distributions}
\end{figure*}

\section{Results}
\vspace{-8pt}
\label{sec:results}
In the pursuit of a method that can act as non-intrusive measure of phonetic content preservation in tasks such as \ac{SE} and lip-to-speech, we first conduct an analysis of correlations of different \acp{SpeechLM} with \ac{PER}. Although perplexity does not have an upper bound and a fixed common range across all methods, its trends provide good indication of each method treats different kinds of input data. \Cref{tab:correlation} shows the performance of the methods presented in \cref{sec:implementation}, measured by Pearson's and Spearman's rank correlation coefficients (\acs{PCC} and \acs{SRCC}) to evaluate linear and ordinal correlation, respectively. We evaluate the methods on noisy mixtures and on speech generated from silent videos. In noisy conditions, the supervised Squim STOI has the best correlation with \ac{PER}, but interestingly, SpeechGPT has good performance, rivaling that of DistillMOS. For the case of lip-to-speech data, TWIST is the most correlated. While methods like TWIST and SpeechGPT remain relatively stable across the two datasets, DistillMOS and Squim STOI have low correlations in lip-to-speech data.

When comparing SpeechLMScore with layers 3 and 6, it is apparent that layer 6 is the most adequate for measuring linguistic content, with higher correlations on both sets. Another finding is that the combination of an \ac{ASR} system with an \ac{LLM} performs much better in this task when a less capable \ac{ASR} model is used. The system using Parakeet had lower correlation with \ac{PER} than the one using QuartzNet. This can be explained by Parakeet's robustness to noise, which results in generally more cohesive inputs for the \ac{LLM} model further down the pipeline.

To assess the models' sensitiveness to the case of complete gibberish, we compare the distributions of the values obtained for gibberish speech, the original LRS3 test set
and the noisy mixtures at -5dB \ac{SNR}. \cref{fig:distributions} shows the densities for the different methods. SpeechLMScore, DistillMOS and UTMOSv2 do not seem to be capable of distinguishing gibberish from the original audio. On the other hand, TWIST and the \ac{ASR} + \ac{LLM} pipeline are the best at discerning these cases, with different behaviors for the distribution of noisy data.

Finally, we report the correlations with \ac{WER} obtained from a listening experiment. Ten annotators were asked to transcribe audio files generated from different lip-to-speech methods compared in \cite{deoliveira2025lipdiffuser}. In total, there are 125 files, with 2 transcriptions per file, whose scores are averaged. In this experiment, the QuartzNet + GPT-2 framework has the best correlations, with TWIST as a close second. As in \cref{tab:correlation}, VAE-GSLM also has good correlation on lip-to-speech data.

\begin{table}
    \centering
    \caption{Correlations with \ac{WER} from human annotators on lip-to-speech data, generated by multiple systems.}%
    \begin{tabular}{lccc}
        \toprule
         && $|\mathrm{PCC}|$ & $|\mathrm{SRCC}|$\\
        \midrule
        SpeechLMScore (3) && 0.321 & 0.281\\
        SpeechLMScore (6) && 0.438 & 0.377\\
        VAE-GSLM          && 0.550 & 0.503\\
        TWIST             && 0.582 & 0.530\\
        SpeechGPT         && 0.395 & 0.312\\
        QuartzNet + GPT-2  && \textbf{0.593} & \textbf{0.639}\\
        Parakeet + GPT-2   && 0.496 & 0.508\\
        \midrule
        UTMOSv2           && 0.150 & 0.133\\
        DistillMOS        && 0.012 & 0.015\\
        \bottomrule
    \end{tabular}%
    \label{tab:listening}
\end{table}

\section{Conclusion}
\label{sec:conclusion}
\vspace{-10pt}
We conducted a study on log-perplexity as a measure of gibberishness in audio, adapting various \acp{SpeechLM} for this task. We found SpeechLMScore to have better performance when using discrete tokens from the 6th layer rather than the 3rd, although it still fails to make a clear distinction between meaningful and gibberish speech. In a conventional noisy mixture setting, SpeechGPT was the best in correlation with the intrusive \ac{PER}. Overall, the most balanced method was TWIST, with consistent performance across all experiments. Finally, we published a test set of high-quality synthesized gibberish samples for further development of methods for assessment of gibberishness in speech.

\clearpage
\section{References}\vspace{-6px}
\label{sec:refs}
\atColsBreak{\vskip5pt}
\printbibliography[heading=none]

\end{document}